\documentclass[a4paper,twocolumn,english,pre]{revtex4}
\usepackage[T1]{fontenc}
\usepackage[latin1]{inputenc}
\usepackage{amsmath}
\usepackage{graphicx}
\usepackage{amssymb}

\makeatletter

\providecommand{\LyX}{L\kern-.1667em\lower.25em\hbox{Y}\kern-.125emX\@}

\usepackage{amsfonts}
\usepackage{amsmath}
\usepackage{geometry}
\usepackage{graphicx}

\usepackage{babel}
\makeatother
\begin{document}

\title{%
\marginpar{%
}Statistics of transmitted power in multichannel dissipative ergodic
structures}

\author{Igor Rozhkov$^{1}$, Yan V. Fyodorov$^{2,3}$, Richard L. Weaver$^{1}$}

\address{$^{1}$Department of Theoretical and Applied Mechanics, University
of Illinois at Urbana-Champaign, Urbana, IL 61801}

\address{$^{2}$Department of Mathematical Sciences, Brunel University, Uxbridge
UB83PH, United Kingdom }

\address{$^{3}$Petersburg Nuclear Physics Institute, Gatchina 188350, Russia}

\begin{abstract}
We use the Random Matrix Theory (RMT) to study the probability distribution
function and moments of the wave power transmitted inside systems
with ergodic wave motion. The results describe either open multichannel
systems or their closed counterparts with local-in-space internal
dissipation. We concentrate on the regime of broken time-reversal
invariance and employ two different analytical approaches: the exact
supersymmetry method and a simpler technique that uses RMT eigenstatistics
for closed non-dissipative systems as an input. The results of the
supersymmetric method were confirmed by numerical simulation. The
simpler method is found to be adequate for closed systems with uniform
dissipation, or in the limit of a large number of weak local dampers. 
\end{abstract}
\maketitle

\section{Introduction}

Transport through open chaotic systems is often viewed as a scattering
process. Standard examples of systems of that kind are compound-nuclei,
mesoscopic quantum dots and wires, microwave cavities and acoustic
or ultrasonic bodies {[}1-12{]}. Incident waves are introduced into
a disordered or irregularly shaped part of the structure via channels,
e.g. waveguides or infinitely long ideal leads. Assuming negligible
dissipation, transport properties are obtained by relating incident
and outgoing wave fields in terms of the unitary scattering matrix
$S$. If internal dissipation is not negligible, it can be simulated
by the action of additional open channels \cite{key-9}. This is the
case, for example, with microwave cavities where non-perfectly reflecting
walls cause loss of wave energy or with ultrasonic solids where internal
friction acts in the bulk \cite{key-8,key-9,key-12}.

The scattering model thus applies to open systems both with and without
internal dissipation. The scattering approach provides a useful tool
for statistical characterization of chaotic transport. Assuming the
wave dynamics inside the system to be ergodic so that the entire phase
space of the system is explored {[}2-6{]}, the scattering approach
is combined with a statistical analysis based on Random Matrix Theory.
For systems without losses, one can make assumptions on the statistics
of the $S$-matrix \cite{key-2,key-11}. An alternative method uses
a Random Matrix assumption on the level of the wave equation associated
with the closed non-dissipative structure. Here the basic object is
the Green function (resolvent) related to that wave equation, and
the method works equally well in both open and closed, dissipative
and non-dissipative complex structures {[}1-9,10,12{]}. Matrices from
a Random Matrix Ensemble then replace the wave equation's linear differential
operator, and the problem of constructing various moments of the transport
characteristics is expressed in terms of ensemble averages of the
products of the resolvents. Transport characteristics calculated in
that way are known, under certain conditions, to describe results
of experimental measurements in systems with ergodic wave motion {[}1-6{]}.

In this paper we are interested in characterizing the wave power $T$
transmitted between a source at site $i$ and a receiver at site $j$
in a closed system with internal losses. The statistics of $T$ are
potentially useful for studies of power transmission in complex reverberant
structures {[}12-14{]}, where both mean power and the magnitude of
its fluctuations away from the mean are important. In particular,
we wish to calculate average $T$ and $T^{2}$ with the ultimate goal
to compare with measurements such as these of Ref. {[}12{]}. The complex
amplitude of the transmitted wave is simply proportional to the off-diagonal
matrix element of the resolvent: $G(E)\equiv $ $\left[E\, I+i\varepsilon \, I-H+i\Gamma \right]^{-1}$.
Here $H$ corresponds to the Hamiltonian of the closed non-dissipative
chaotic structure. The matrix $\Gamma $ describes coupling to external
channels or internal local-in-space losses, $I$ is the identity matrix,
the parameter $\varepsilon >0$ describes uniform dissipation and
$E$ is the spectral variable. The quantity of prime interest is $T=\left|G_{ij}\left(E\right)\right|^{2},i\neq j$,
i.e. the product of retarded and advanced Greens functions (propagators):
$G_{ji}^{R}(E)\equiv $ $\left[E\, I+i\varepsilon \, I-H+i\Gamma \right]_{ji}^{-1}$
and $G_{ij}^{A}=\left(G_{ji}^{R}(E)\right)^{\ast }$ respectively.\ Except
for slowly varying factors of receiver gain and source strength, the
quantity $T$ represents the ultrasonic power of Ref. {[}12{]}; see
also {[}10,13{]}.

The fluctuations in $T$, as measured in Ref. {[}12{]} were in only
modest agreement with theoretical predictions based on a simplified
version of the random matrix approach. The moments of $T$ were calculated
there using a naive form of ensemble averaging, {[}8,12-14{]}. This
relatively simple approach uses statistical assumptions for eigenfunctions
and the real parts of the eigenvalues of the open (dissipative) system
identical to those of the corresponding closed system. As will be
seen later, such an assumption is strictly justified only for a special
case of uniform dissipation. In a more general situation this approach
fails. A proper treatment calls for a more elaborate technique, which
we outline and present below.

When losses are negligible the systems discussed in Ref. {[}12-14{]}
are invariant under time reversal. The appropriate choice for the
corresponding random matrix $H$ should therefore be a real symmetric
matrix taken e.g. from the Gaussian Orthogonal Ensemble (GOE). In
principle, the powerful methods of ensemble averaging we employ here
can be used for such an ensemble, but the calculations are technically
involved and will be presented in a separate publication.

Here we address ourselves to the somewhat simpler case in which $H$
is complex Hermitian, generic for systems with broken time reversal
invariance. Correspondingly, $H$ is treated as a Hermitian $N\times N$
matrix consisting of uncorrelated centered random complex numbers,
their variances defined by: $\left\langle H_{ij}H_{lm}^{\ast }\right\rangle =\left(\lambda ^{2}/N\right)\delta _{il}\delta _{jm}$
with angular brackets indicating ensemble averaging. Such an $H$
is a member of the Gaussian Unitary Ensemble (GUE) of random matrices
{[}1{]}. Although our present results on the statistics of $T$ are
not directly applicable to the time-reversal invariant systems discussed
in {[}12-14{]}, they may elucidate the discrepancies found in Ref.
{[}12{]} between measurements and the predictions of the 'naive' averaging.
They also develop and illustrate the mathematical methods which will
be used for a proper non-perturbative analysis of the time reversal-invariant
problem. The present calculations are also relevant for scattering
systems with broken time reversal invariance as exemplified in certain
chaotic billiards {[}8{]}, optical and semiconductor superlattices
{[}15{]} and quantum graphs {[}16{]}. In fact, our results on the
distribution of the \textit{off-diagonal} elements of the resolvent
extend earlier studies concentrated on \textit{diagonal} entries for
the same quantity, see {[}22-23{]} and references therein. Let us
finally mention that there exist clear analogy between our research
and that presented in the paper {[}19{]}, see also the review {[}20{]}.
However, the model considered in {[}19-20{]} did not take local dampers
into account, but rather addressed effects of Anderson localization.

The damping matrix $\Gamma $ is in general Hermitian positive semi-definite.
In our model there is no loss of generality in assuming it to be diagonal.
Indeed, in view of the rotational invariance of the Gaussian Unitary
ensemble: $H\mapsto UHU^{-1}$ ($U^{-1}=U^{\dag }$) we always can
select the basis which diagonalizes $\Gamma $, bringing it to the
form $\Gamma =diag\{\gamma ,\gamma ,...\gamma ,0,...0\}$. The number
$M<N$ of non-zero entries can be interpreted either as a number of
equivalent open channels in the scattering system {[}3,4,9{]} or a
number of equivalent localized 'dampers' in a closed system with losses.
While we take all the $\gamma $s to be equal, the expressions we
develop are easily generalized to the case of varying damper strengths.
It should be stressed that in general the matrices $\Gamma $ and
$H$ do not commute, and therefore the eigenvectors and eigenvalues
of the 'effective non-Hermitian Hamiltonian' $H-i\Gamma $ are not
trivially related to those of $H$. This very fact makes the naive
averaging incorrect. In contrast, the term $i\varepsilon \, I$ interpreted
as the 'uniform damping' preserves eigenvectors of $H$ and just adds
a uniform shift $i\varepsilon $ to all eigenvalues.

The presence of an $N\times N$ random Hamiltonian $H$ in the expression
for the resolvent matrix $G$ enables us to carry out the ensemble
averaging exactly using the supersymmetry method {[}3,4,17,18,20,21{]}.
Application of this non-perturbative technique leads to an expression
for the entire probability distribution function of $T$. We present
the corresponding derivation in Section II. In Section III, we compare
the results for the first two moments of $T$ as obtained using the
supersymmetry method and the methods of Ref. {[}12-14{]}. The results
are then verified numerically by direct simulation of the model. Section
IV contains conclusions.

\section{Probability distribution function of the power. Supersymmetric calculation}

In the previous Section we defined the power $T$ as a product of
advanced and retarded Green functions $G_{ij}$. Our goal is to compute
the statistics, i.e. ensemble averages: $\left\langle T\right\rangle _{H}$,
$\left\langle T^{2}\right\rangle _{H}$, etc., where subscript $H$
designates averaging with the Gaussian weight $\exp \left\{ -\frac{N}{2\lambda ^{2}}TrH^{\dagger }H\right\} $.
At the first stage of the supersymmetric calculation we make use of
the following identities for the inverse propagator $D_{ij}=\left[E+i\varepsilon -H+i\Gamma \right]_{ij}$
{[}3,4,17,18,20,21{]}: \begin{gather*}
\det \mathfrak{D}_{b}^{-1}=\int \left[ dS^{\dagger }\right] \left[ dS\right]
\exp \left\{ i\mathfrak{L}_{b}\mathfrak{\,}\left( E,S\right) \right\} , \\
\det \mathfrak{D}_{f}=\left( -1\right) ^{N}\int \left[ d\chi ^{\ast }\right] \left[ d\chi \right] \exp \left\{ i\mathfrak{L}_{f}\mathfrak{\,}\left(
E,\chi \right) \right\} .
\end{gather*}Here we introduced $2N-$dimensional vectors $S^{T}=\left(S_{1}^{T},S_{2}^{T}\right)$
and $\chi ^{T}=\left(\chi _{1}^{T},\chi _{2}^{T}\right)$, consisting
of complex commuting or bosonic (b) variables and anticommuting or
fermionic (f) variables respectively. $\mathfrak{D}_{b}=diag\left\{ D,-D^{\dagger }\right\} $
and $\mathfrak{D}_{f}=diag\left\{ D,D^{\dagger }\right\} $ are $2N\times 2N$
block diagonal matrices, and $\mathfrak{L}_{b}\, \left(E,S\right)=S^{\dagger }\mathfrak{D}_{b}S\, $,
$\mathfrak{L}_{f}\, \left(E,\chi \right)=\chi ^{\dagger }\mathfrak{D}_{f}\chi \, $.
The negative sign of $\mathfrak{D}_{b\, \, }^{22}$ is necessary for
convergence of the integrals in what follows. Differentiating the
first equality with respect to $\mathfrak{D}_{b\, \, ji}^{11}$ and
$\mathfrak{D}_{b\, \, ij}^{22}$, and then combining the result with
the second equality, we obtain: \begin{alignat}{2}
T= & \mathfrak{D}_{ij}^{-1\, }\mathfrak{D}_{\, \, \, \, i\, j}^{\ast -1}\notag  &  & \\
= & \int \left[d\Phi ^{\dagger }\right]\left[d\Phi \right]S_{1j}^{\ast }S_{1i}S_{2i}^{\ast }S_{2j}\exp \left\{ i\mathfrak{L}\, \left(E,\Phi \right)\right\} , &  & 
\end{alignat}
where the integration involves four-component supervectors $\Phi ^{T}=\left(S^{T},\chi ^{T}\right)$,
and where $\mathfrak{L}\, \left(E,S\right)=\mathfrak{L}_{b}\, \left(E,S\right)+\mathfrak{L}_{f}\, \left(E,\chi \right)=\Phi ^{\dagger }\mathfrak{D}\Phi $,
$\mathfrak{D}=diag\left\{ \mathfrak{D}_{b},\mathfrak{D}_{f}\right\} $
{[}3,4,17,18,20,21{]}. Because the random matrix $H$ is in the exponent,
$T$ is now suitable for ensemble averaging. 

In a similar fashion, employing the Wick theorem one can verify the
following formula necessary for the calculation of an arbitrary moment
of transmitted power $\left\langle T^{n}\right\rangle _{H}$: \begin{align}
T^{n}&=\left( n!\right) ^{-2}\int \left[ d\Phi ^{\dagger }\right] \left[
d\Phi \right] S_{1j}^{\ast n}S_{1i}^{n}S_{2i}^{\ast n}S_{2j}^{n} \notag \label{Tn} \\
\times & \exp \left\{
i\mathfrak{L\,}\left( E,\Phi \right) \right\}
=\left\langle S_{1j}^{\ast n}S_{1i}^{n}S_{2i}^{\ast
n}S_{2j}^{n}\right\rangle _{\Phi }.
\end{align}A shorthand notation $\left\langle ...\right\rangle _{\Phi }$ has
been introduced for the 'Gaussian' integration over the supervector
components. Hereafter we use the more convenient '{[}1,2{]}' ('retarded-advanced')-block
notation for supervectors and supermatrices, see for example Ref.
{[}4{]}. With the supervector $\Psi ^{T}=\left(S_{1}^{T},\chi _{1}^{T},S_{2}^{T},\chi _{2}^{T}\right)$
and the $4\times 4$ supermatrices $L=diag\left\{ 1,1,-1,1\right\} $,
$\Lambda =diag\left\{ 1,1,-1,-1\right\} $ the exponent in the integrand
reads:\begin{gather*}
\mathfrak{L\,}\left( E,\Psi \right) =E\Psi ^{\dagger }\left( I\otimes
L\right) \Psi +i\Psi ^{\dagger }\left( \Gamma \otimes \Lambda L\right) \Psi \\
-\Psi ^{\dagger }\left( H\otimes L\right) \Psi +i\varepsilon \Psi ^{\dagger
}\left( I\otimes L\right) \Psi ,
\end{gather*}and, Eq. (\ref{Tn}) becomes:\begin{equation}
T^{n}=\left(n!\right)^{-2}\left\langle F^{n}\left[\Psi \right]\right\rangle _{\Psi },F\left[\Psi \right]=S_{1j}^{\ast }S_{1i}S_{2i}^{\ast }S_{2j}.\label{Tn1}\end{equation}

The ensemble averaging can now be performed with the aid of the identities
{[}4{]}:\begin{gather*}
\left\langle \exp \left\{ i\Psi ^{\dagger }\left( H\otimes L\right) \Psi
\right\} \right\rangle _{H}=\exp \left\{ -\frac{1}{2N}StrA^{2}\right\} , \\
\quad A_{pq}^{\left( kl\right) }=L_{kk,pp}^{1/2}\sum\limits_{i=1}^{N}\left(
\Psi _{i}\right) _{k}^{p}\left( \Psi _{i}^{\dagger }\right)
_{l}^{q}L_{ll,qq}^{1/2}, \\
\quad StrAL=\Psi ^{\dagger }\left( I\otimes L^{1/2}\Lambda L^{1/2}\right)
\Psi =\Psi ^{\dagger }\left( I\otimes \Lambda L\right) \Psi ,
\end{gather*}where we have set $\lambda =1$, and introduced a $4\times 4$ supermatrix
$A$. Thus\begin{widetext}

\begin{equation}
\left\langle T^{n}\right\rangle _{H} =\left( n!\right) ^{-2}\int \left[d\Psi ^{\dagger }\right] \left[ d\Psi \right] F^{n}\left[ \Psi \right] \exp \left\{ iE\Psi ^{\dagger }\left( I\otimes L\right) \Psi -\Psi
^{\dagger }\left( \Gamma \otimes \Lambda L\right) \Psi -\frac{1}{2N}StrA^{2}-\varepsilon StrAL\right\} ,  
\end{equation}

\end{widetext}where $k$, $l$ distinguish between retarded and advanced, ($1$
and $2$) supermatrix blocks indices and $p$, $q$ equal to $b$
or $f$ {[}4,17{]}. The next stages of supersymmetric procedure include
{[}1,3,4,17,18,20,21{]}: 1) the Hubbard-Stratonovich transformation,
that removes quartic (in $\Psi $) term in the exponential; 2) $\Psi -$variables
integration; 3) evaluation of the remaining integral using saddle
point approximation in the limit $N\rightarrow \infty $. We have,
after step 1):\begin{widetext}

\begin{align}
\left\langle T^{n}\right\rangle _{H}  =\left( n!\right) ^{-2}& \int \left[ dR\right] \exp \left\{ -\frac{N}{2}StrR^{2}+i\varepsilon NStrR\Lambda
+iStrRA\right\} \\
\times & \int \left[ d\Psi ^{\dagger }\right] \left[ d\Psi \right] F^{n}\left[ \Psi \right] \exp \left\{ i\left( E\Psi ^{\dagger }L\Psi +i\Psi
^{\dagger }\left( \Gamma \otimes \Lambda L\right) \Psi \right) \right\} . 
\notag
\end{align}

\end{widetext}

Since $StrRA=\Psi ^{\dagger }L^{1/2}RL^{1/2}\Psi $, for an arbitrary
$4\times 4$ supermatrix $R$,\begin{widetext}

\begin{align}
\left\langle T^{n}\right\rangle _{H}=\left( n!\right) ^{-2} &\int \left[ dR\right] \exp \left\{ -\frac{N}{2}StrR^{2}+i\varepsilon NStrR\Lambda \right\} 
\notag \int \left[ d\Psi ^{\dagger }\right] \left[ d\Psi \right] F^{n}\left[ \Psi \right] \label{ttnh} \\
\times & \exp \left\{ -i\Psi ^{\dagger }L^{1/2}\left( -EI_{4}\otimes
I_{N}-R\otimes I_{N}-i\Lambda \otimes \Gamma \right) L^{1/2}\Psi \right\} . 
\end{align}

\end{widetext}

Using the Gaussian nature of the integral: \begin{equation*}
\int \left[ d\Psi ^{\dagger }\right] \left[ d\Psi \right] \exp \left\{
-i\Psi ^{\dagger }f\Psi \right\} =S\det f^{-1},
\end{equation*}the following general relation, can be derived similarly to Eq. (\ref{Tn1}):\begin{align}
&\int \left[ d\Psi ^{\dagger }\right] \left[ d\Psi \right] \left[ \left(
\Psi _{i}\right) ^{b}\left( \Psi _{j}^{\dagger }\right) ^{b}\left( \Psi
_{i}\right) ^{b}\left( \Psi _{j}^{\dagger }\right) ^{b}\right] ^{n} \label{wt} \\
\times&\exp
\left\{ -i\Psi ^{\dagger }f\Psi \right\} =\left( n!\right) ^{2}f_{\quad 12,bb_{ij}}^{-1}f_{\quad
21,bb_{ji}}^{-1}S\det f^{-1}  \notag
\end{align}

Setting $f=L^{1/2}\left(-EI_{4}\otimes I-R\otimes I-i\Lambda \otimes \Gamma \right)L^{1/2}$,
we integrate out the components of $\Psi $ with the help of Eq. (\ref{wt}):\begin{equation}
\left\langle T^{n}\right\rangle _{H}=\int \left[dR\right]F^{n}\left[\mathfrak{G}\right]\exp \left\{ -N\mathcal{L}\left[R\right]+\delta \mathcal{L}\right\} ,\label{Tn2}\end{equation}
where\begin{align*}
F\left[ \mathfrak{G}\right] & =\mathfrak{G}_{\ 12,\,bb\,}^{-1}\mathfrak{G}_{\ 21,bb}^{-1},\mathfrak{G}=-EI_{4}-R, \\
\mathcal{L}\left[ R\right] & =\frac{1}{2}StrR^{2}+Str\ln \left(
-EI_{4}-R\right), \\
\delta \mathcal{L}=& i\varepsilon N\,StrR\Lambda \\
&-M\,Str\ln \left[
I_{4}-i\gamma \Lambda \left( -EI_{4}-R\right)^{-1}\right].
\end{align*}See Appendix A for the details. 

$\left\langle T^{n}\right\rangle _{H}$ is now written as an integral
over $4\times 4$ supermatrix $R$. The stationarity condition for
$\mathcal{L}\left[R\right]$, in the limit of large $N$, yields a
stationary point $R_{s}$, satisfying: $R_{s}=1/\left(-EI_{4}-R_{s}\right)$.
The solution is not unique, it is a saddle manifold in a space of
$4\times 4$ supermatrices, spanned by 'pseudounitary' supermatrices
$\mathfrak{T}$:$\, \, R_{s}=-EI_{4}/2+i\pi \nu \mathfrak{T}^{-1}\Lambda \mathfrak{T}=-EI_{4}/2-\pi \nu Q$,
where $\nu =\sqrt{4-E^{2}}/\left(2\pi \right)$ is Wigner's semicircular
mean density of eigenvalues (GUE, $\lambda =1)$. See Refs. {[}4,17{]}
for the explicit form of supermatrix $Q$. 

After integrating out local fluctuations over directions $R$ orthogonal
to the manifold of stationary points (the procedure is asymptotically
exact for large $N$) the remaining integration goes over the manifold
parametrized by $Q$:\begin{align}
\left\langle T^{n}\right\rangle _{H}& =\left( \pi \nu \right) ^{2n}\int 
\left[ dQ\right] \left( Q_{12,bb}Q_{21,bb}\right) ^{n}  \notag \\
& \times S\det \,^{-M}\left[ I_{4}+i\frac{E}{2}\gamma \Lambda +i\pi \nu
\gamma Q\Lambda \right] \notag \\
&\times \exp \left\{ -i\varepsilon \pi \nu NStrQ\Lambda
\right\} .
\end{align}

The expression for the $n$-th moment of power allows one to find
the entire distribution function $P\left(T\right)$, cf. {[}17,19,20{]}:\begin{equation*}
\mathcal{P}\left( T\right) =\int \left[ dQ\right] \delta \left( T-\left( \pi
\nu \right) ^{2}Q_{12,bb}Q_{21,bb}\right) Y\left( Q\right) ,
\end{equation*}or, for the 'scaled' power $y=T/\left(\pi \nu \right)^{2}$:\begin{equation}
\mathcal{P}\left(y\right)=\int \left[dQ\right]\delta \left(y-Q_{12,bb}Q_{21,bb}\right)Y\left(Q\right),\label{Py1}\end{equation}
where\begin{align}
Y\left(Q\right) & =S\det \, ^{-M}\left[I_{4}+i\frac{E}{2}\gamma \Lambda +i\pi \nu \gamma Q\Lambda \right]\notag \\
\times  & \exp \left\{ -i\varepsilon \pi \nu NStrQ\Lambda \right\} \notag .
\end{align}
Evaluation of the superintegral in Eq. (\ref{Py1}) is presented in
Appendix A. The result reduces to:\begin{widetext}

\begin{gather}
\mathcal{P}\left( y\right) =\delta \left( y\right) +\left( \frac{d}{dy}+y\frac{d^{2}}{dy^{2}}\right) \int_{1}^{\infty }d\lambda
_{1}\int_{-1}^{1}d\lambda _{2}\delta \left( y+1-\lambda _{1}^{2}\right)  
\frac{\lambda _{1}^{2}-\lambda _{2}^{2}}{\left( \lambda _{1}-\lambda
_{2}\right) ^{2}}\exp \left\{ -\epsilon \left( \lambda _{1}-\lambda
_{2}\right) \right\} \left( \frac{g+\lambda _{2}}{g+\lambda _{1}}\right) ^{M}
\label{Py2}  \\
=\left( \frac{d}{dy}+y\frac{d^{2}}{dy^{2}}\right) \frac{\exp \left\{
-\epsilon \sqrt{1+y}\right\} }{2\sqrt{1+y}\left( g+\sqrt{1+y}\right) ^{M}}  \int_{-1}^{1}d\lambda _{2}\frac{\sqrt{1+y}+\lambda _{2}}{\sqrt{1+y}-\lambda _{2}}\exp \left\{ \epsilon \lambda _{2}\right\} \left( g+\lambda
_{2}\right) ^{M},  \notag
\end{gather}

\end{widetext}where $g=\left(1/\gamma +\gamma \right)/\left(2\pi \nu \right)\, \, $and
$\epsilon =2\pi \nu N\varepsilon $. 

Setting the 'uniform damping' $\epsilon $ to zero we were able to
evaluate the remaining integral explicitly (see Appendix A for\ the
details) and Eq. (\ref{Py2}) yielded:\begin{widetext}

\begin{align}
\mathcal{P}\left( y\right) & =\frac{\left\{ \left( g-1\right)
^{M+1}\,-\left( g+1\right) ^{M+1}\right\} p_{1}+\left\{ \left( g+1\right)
^{M+1}\,+\left( g-1\right) ^{M+1}\right\} p_{2}}{8\sqrt{\left( 1+y\right)
^{5}}\left( g+\sqrt{1+y}\right) ^{M+2}},  \notag \\
p_{1}& =\frac{g\left( y-2\right) }{\left( M+1\right) }\left( g+\left(
M+2\right) \sqrt{1+y}\right) -\left( y+1\right) \left( 2+\left( M+1\right)
\left( y+2\right) \right) , \label{Py3} \\
p_{2}& =2 \left( y+1\right) \left( g+\left( M+2\right) 
\sqrt{1+y}\right) .\notag
\end{align}

\end{widetext} Eqs. (\ref{Py2}) and (\ref{Py3}) constitute the main result of
this Section. 

At this point it is interesting to observe that Eq. (\ref{Py3}) can
be in fact used to cover the case of uniform damping in closed system:
$\epsilon >0,\, M=0$. For this we note that in the limit $\gamma \rightarrow 0$
(i.e. $g\sim 1/2\pi \nu \gamma \rightarrow \infty $) and $M\rightarrow \infty $
but $\gamma M\sim const$ the factor $\left(g+\lambda _{2}\right)^{M}/\left(g+\lambda _{1}\right)^{M}$
in the integrand of Eq. (\ref{Py2}) is converted to $\exp \left\{ 2\pi \nu \gamma M\left(\lambda _{2}-\lambda _{1}\right)\right\} $.
Such a replacement is equivalent to generating an effective uniform
damping $\epsilon =2\pi \nu \gamma M$. The fact that the large number
of weakly open channels (or weak local-in-space dampers) is essentially
equivalent to uniform damping is well known, see e.g. {[}4{]}. Performing
the limit $g\rightarrow \infty ,M\rightarrow \infty $ when keeping
the product $2\pi \nu \gamma M=\epsilon $ finite we find that the
distribution Eq. (\ref{Py3}) is reduced to:\begin{align}
\mathcal{P}\left( y\right) & =\frac{\exp \left\{ -\epsilon \sqrt{1+y}\right\} \sinh \epsilon}{4\epsilon\sqrt{\left( 1+y\right) ^{5}}} \notag \\
&\times\left[\epsilon^{2}\left( y+1\right)\left( y+2\right) -\left( y-2\right)\left( 1+\epsilon \sqrt{1+y}\right)\right] \notag \\
& +\frac{\exp \left\{ -\epsilon \sqrt{1+y}\right\} \cosh \epsilon \left( 1+\epsilon \sqrt{1+y}\right)}{2\sqrt{\left( 1+y\right) ^{3}}} , \label{Pas} 
\end{align}This distribution of transmitted power for systems with uniform damping
is interesting and important on its own.

Let us consider now a few other regimes. For the weakly damped system
($M$ is fixed, $g\gg 1$), Eq. (\ref{Py3}) can be approximated by:\begin{equation}
\mathcal{P}\left(y\right)\sim \frac{4+y}{4\sqrt{\left(1+y\right)^{5}}}+O\left(\frac{1}{g^{2}}\right).\label{Py4}\end{equation}

The asymptotic behavior of $\mathcal{P}\left(y\right)$ in the limit
$y\rightarrow \infty $ for any $M$ and $g$ is given by:\begin{eqnarray}
\mathcal{P}\left(y\right) & \sim  & \frac{\left(M+1\right)}{4y^{\left(M+3\right)/2}}\left\{ \left(g+1\right)^{M+1}-\left(g-1\right)^{M+1}\right\} \notag \\
 &  & +O\left(\frac{1}{y^{\left(M+4\right)/2}}\right),
\end{eqnarray}
which shows that moments $\left\langle y^{n}\right\rangle $ exist
only for $n<\left(M+1\right)/2$. At the same time, as it follows
from Eq. (\ref{Py2}), any non-zero $\epsilon $ guarantees existence
of all moments. For large $y$, the asymptotic forms of the probability
distribution function at non-zero $\epsilon $ are:\begin{equation}
\mathcal{P}\left(y\right)\sim \frac{\epsilon \sinh \epsilon }{\sqrt{y}}\exp \left\{ -\epsilon y^{1/2}\right\} ,\end{equation}
\begin{equation}
\mathcal{P}\left(y\right)\sim \frac{g^{M}\exp \left\{ -\epsilon y^{1/2}\right\} }{y^{\frac{M+1}{2}}},\end{equation}
for $M=0$, and for finite $M$ respectively. 

Finally we want to compare the results of this Section with numerical
solution of the model RMT problem. For this goal we numerically generate
an ensemble of $N\times N$ Hermitian random matrices $[H]$, typically
choosing $1500$ samples from the ensemble and $N=1000$. The entries
of the matrix $H$ are constructed using a random number generator,
with $\left\langle H_{ij}H_{lm}^{\ast }\right\rangle =\left(1/N\right)\delta _{il}\delta _{jm}$.
To simulate the uniform damping and the case of a finite number of
local dampers we take $\Gamma =\varepsilon I$ and $\Gamma =diag\{\gamma ,\gamma ,...,\gamma ,0,...0\}$
(with $M\leq N$ non-zero entries) respectively. Then, for all members
of our ensemble we generate the off-diagonal elements of the resolvent
matrix $G_{ij}(E)=\left[EI+i\Gamma -H\right]^{-1}$ modeling the response
at site $i$ due to excitation at site $j$, with $E$ being the spectral
parameter. 

We first consider the case of uniform damping: $\Gamma =\varepsilon I$.
The modal density $\nu $ for such a system is approximated by Wigner's
semicircular law: $\nu =\sqrt{4-E^{2}}/\left(2\pi \right)$. Therefore,
for a fixed size of the matrices $N$ and spectral variable $E$,
we can explore the range of $\epsilon $ by changing $\varepsilon $.
For $E=0$ the modal density is $\nu =1/\pi $ and so we need not
distinguish between $T$ and $y$. In Fig.1, the numerically obtained
histograms are compared with $\mathcal{P}\left(y\right)$ (Eq. (\ref{Py2}))
for several values of $\epsilon $. We see that numerical results
correspond well with the theoretical curves.%
\begin{figure}[htbp]
\begin{center}\includegraphics{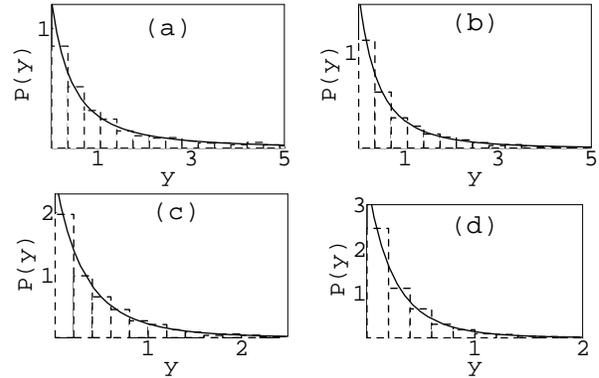}\end{center}

\caption{Probability distribution function (Eq. (\ref{Py2})) and histograms
of power for: (a) $\epsilon =1.0$, (b) $\epsilon =2.0$, (c) $\epsilon =4.0$,
(d) $\epsilon =6.0,$ as obtained numerically. Data are scaled to
unit area. For each plot $1500$ samples of $\left|G_{ij}\left(E\right)\right|^{2},i\neq j$
were computed.}
\end{figure}

This procedure was repeated for the damping matrix $[\Gamma ]=diag\{\gamma ,\gamma ,...,\gamma ,0,...0\}$
with $M$ non-zero entries, by computing $G_{ij}(E=0)$ for different
combinations of parameters $M$ and $g$. We The results are presented
in Fig. 2. Again, the predictions of the supersymmetry method agree
well with the numerical results.

\begin{figure}[htbp]
\begin{center}\includegraphics{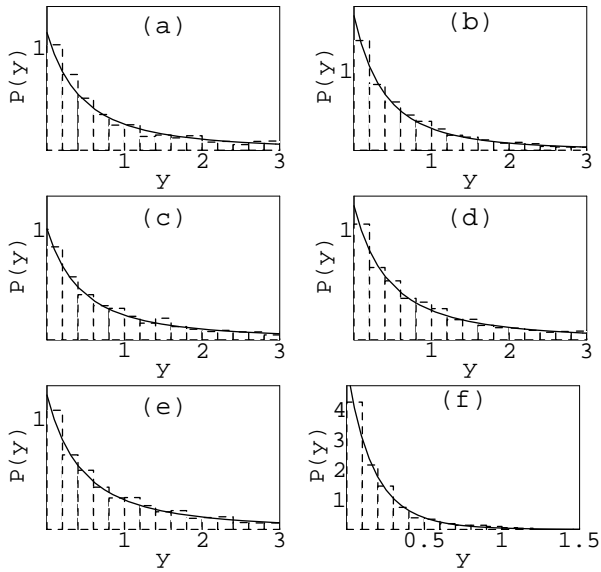}\end{center}

\caption{Probability distribution function (Eq. (\ref{Py3})) and histograms
of power for: (a) $M=2,g=2.16$, (b) $M=6,g=2.16$ (c) $M=40,g=400$,
(d) $M=40,g=40$, (e) $M=400,g=400$, (f) $M=400,g=40$. Data are
scaled to unit area. For each plot $1500$ samples of $\left|G_{ij}\left(E\right)\right|^{2}$
were computed. We imposed the restrictions: $i\neq j$, and $i>M,j>M$
for the non-uniform damping case, to avoid 'recording' the response
from damped sites or from the 'source' site $j$, and to correspond
to the assumptions in the theoretical analyses. Note, that for large
values of $g$ (plots (c) and (e)), $P\left(y\right)$ is not sensitive
to either $g$ or $M$ (Eq. (\ref{Py4})).}
\end{figure}

\section{Moments of power}

In this Section we analyze the first two moments of power $y$ using
two different approaches, both based on the RMT. These moments as
obtained using the supersymmetric calculation (Eqs. (\ref{Py2}),\ (\ref{Py3}))
will be compared to those obtained using the naive approach. We first
consider uniform damping and then $M\neq 0$. 

In the simpler, but inexact, approach $G_{ij}$ is constructed as
a modal sum,\begin{equation*}
G_{ij}\left( E\right) =\sum_{r}\frac{u_{i}^{r}u_{j}^{r\ast }}{E-E_{r}-i\zeta
_{r}},
\end{equation*}and then averaged using the eigenstatistics of the undamped GUE system.
Here $u^{r}$ is the $r$th eigenmode and we call the imaginary part
$\zeta _{r}$ of the eigenenergy $E_{r}$ the resonance width. 

We first consider the case when $\zeta _{r}$ is uniform: $\zeta _{r}=\varepsilon ,$
for all $r$ {[}12{]}:\begin{equation}
y\left(\pi \nu \right)^{2}=\sum _{r}\sum _{m}\frac{u_{i}^{r}u_{j}^{r\ast }}{E-E_{r}-i\varepsilon }\frac{u_{i}^{m\ast }u_{j}^{m}}{E-E_{m}+i\varepsilon }.\label{y1}\end{equation}
On averaging Eq. (\ref{y1}) over the eigenmodes $u^{r}$, and assuming
they are uncorrelated, $\left\langle y\right\rangle $ becomes:\begin{equation*}
\left\langle y\right\rangle \left( \pi \nu \right) ^{2}=\sum_{r}\frac{\left\langle \left\vert u\right\vert ^{2}\right\rangle ^{2}}{\left(
E-E_{r}-i\varepsilon \right) \left( E-E_{r}+i\varepsilon \right) }.
\end{equation*}The summation over the eigenenergies $E_{r}$ is then replaced with
an integral ($\sum _{r}\rightarrow N\nu \int dE_{r}$)\begin{equation*}
\left\langle y\right\rangle =\frac{\left\langle \left\vert u\right\vert
^{2}\right\rangle ^{2}}{\left( \pi \nu \right) ^{2}}\int_{-\infty }^{+\infty
}\frac{\left( N\nu \right) dE_{r}}{\left( E-E_{r}-i\varepsilon \right)
\left( E-E_{r}+i\varepsilon \right) },
\end{equation*}where $\nu =\sqrt{4-E^{2}}/\left(2\pi \right)$ is the GUE modal density.
Therefore, in the uniform damping case, the naive procedure produces:\begin{align}
\left\langle y\right\rangle & =\pi \nu \frac{\left\langle \left\vert
u\right\vert ^{2}\right\rangle ^{2}}{\left( \pi \nu \right) ^{2}}\frac{\pi }{\varepsilon } \label{yb1} \\
& =N^{2}\left\langle \left\vert u\right\vert ^{2}\right\rangle ^{2}\frac{2}{2\pi \nu N\varepsilon }=\frac{2}{\epsilon },\quad   \notag
\end{align}where $\left\langle \left|u\right|^{2}\right\rangle $ has been set
to $1/N$ (by normalization). 

The second moment of power is calculated in Appendix B by means of
the same approach, and is given by:\begin{align}
\label{ysb1} \left\langle y^{2}\right\rangle & =\frac{1}{\pi ^{3}\nu ^{3}N^{3}\varepsilon
^{3}}+\frac{1}{4\pi ^{4}\nu ^{4}N^{4}\varepsilon ^{4}} \\
& \times \left[ 1+8\pi ^{2}\left( N\nu \right) ^{2}\varepsilon ^{2}-\exp
\left( -4\pi N\nu \varepsilon \right) \right]   \notag
\end{align}

For the uniform damping case ($\zeta _{r}=\varepsilon $ for all modes),
application of the results of Section II is especially straightforward.
As already discussed, Eq. (\ref{Py2}) shows that in our model this
case is realized either by setting $M=0$ with finite $\epsilon $
or by letting $M$ be large and $\gamma $ be small, such that $\gamma M\, \, $is
finite. One can use Eq. (\ref{Pas}) for this purpose, but it is more
convenient to start with the first part of Eq. (\ref{Py2}). Setting
$M=0$ and introducing:\begin{align*}
\mathfrak{f=}&\int_{1}^{\infty }d\lambda _{1}\int_{-1}^{1}d\lambda _{2}\delta
\left( y+1-\lambda _{1}^{2}\right) \\ 
&\times \frac{\lambda _{1}^{2}-\lambda _{2}^{2}}{\left( \lambda _{1}-\lambda _{2}\right) ^{2}}\exp \left\{ -\epsilon \left(
\lambda _{1}-\lambda _{2}\right) \right\} ,
\end{align*}we integrate by parts in Eq. (\ref{Py2}):\begin{gather}
\mathcal{P}\left( y\right) =\delta \left( y\right) +\left( \frac{d}{dy}+y\frac{d^{2}}{dy^{2}}\right) \mathfrak{f,}  \notag \\
\left\langle y^{n}\right\rangle =\int_{0}^{\infty }y^{n}\mathcal{P}\left(
y\right) dy=-y^{n}\left( n\mathfrak{f}+y\mathfrak{f}_{y}^{\prime }\right)
\mid _{0}^{\infty } \label{Pofy1} \\
+n^{2}\int_{0}^{\infty }y^{n-1}\mathfrak{f}dy=n^{2}\int_{0}^{\infty }y^{n-1}\mathfrak{f}dy.  \notag
\end{gather}Integration with respect to $y$ in Eq. (\ref{Pofy1}) eliminates
the delta function, and gives:\begin{eqnarray}
\left\langle y\right\rangle  & = & \int _{1}^{\infty }d\lambda _{1}\int _{-1}^{1}d\lambda _{2}\exp \left\{ -\epsilon \left(\lambda _{1}-\lambda _{2}\right)\right\} \notag \\
 & \times  & \left(\frac{\lambda _{1}+\lambda _{2}}{\lambda _{1}-\lambda _{2}}\right)=\frac{2}{\epsilon },\label{yb2}
\end{eqnarray}
\begin{eqnarray}
\left\langle y^{2}\right\rangle  & = & 4\int _{0}^{\infty }y\mathfrak{F}dy=\frac{4}{\epsilon ^{4}}\left(1-\exp \left\{ -2\epsilon \right\} \right)\notag \\
 &  & +\frac{8}{\epsilon ^{2}}\left(1+\frac{1}{\epsilon }\right),\label{yb3}
\end{eqnarray}
which takes the same form as Eqs. (\ref{yb1}) and (\ref{ysb1}) upon
substitution of $2\pi \nu N\varepsilon $ for $\epsilon $. Thus,
for uniform damping, the 'naive' and supersymmetric methods agree,
for both $\left\langle y\right\rangle $ and $\left\langle y^{2}\right\rangle $.
This is not unexpected, because uniform damping with $M=0$ leaves
eigenstatistics identical to those of closed systems, merely shifting
all eigenenergies by $i\varepsilon $. The results (\ref{yb2}) and
(\ref{yb3}) can readily be reproduced by using $P(y)$ as given by
Eq. (\ref{Pas}). These moments are plotted in Fig. 3 together with
the results of numerical simulations. The first two moments of $y$
were obtained numerically, by inverting matrix $EI+i\Gamma -H$ for
each member of the ensemble. More precisely - we computed the column
vector $G_{ij}(E)$ \ ($j$ is fixed, $i=1,..N$) by solving the
algebraic equations:$[EI+i\Gamma -H]G_{ij}=\delta _{ij}$ for a fixed
value of $E$. Repeating this procedure $1500$ times and averaging
over the ensemble of $H$ and over the $N-1$ values of $i\neq j$,
we obtained $\left\langle y\right\rangle $ and $\left\langle y^{2}\right\rangle $
for $\epsilon =1,2,4,6$. As seen in Fig. 3, the correspondence is
excellent.

\begin{figure}[htbp]
\begin{center}\includegraphics{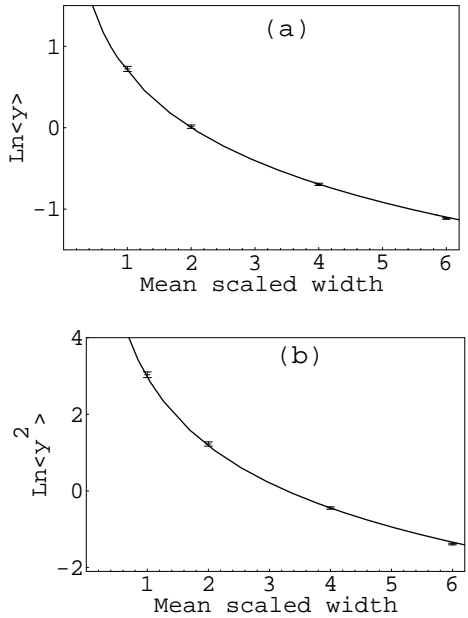}\end{center}

\caption{(a) $Log\left\langle y\right\rangle $ and (b) $Log\left\langle y^{2}\right\rangle $
plotted as the functions of parameter $\epsilon $ for the case of
uniform damping. The solid lines represent theoretical predictions
(Eqs. (\ref{yb1}) and (\ref{ysb1})). The one sigma error bars were
computed based on the observed variances of $y$ and $y^{2}$.}
\end{figure}

Our next goal is to calculate $\left\langle y\right\rangle $ and
$\left\langle y^{2}\right\rangle $ for the system with $M$ equivalent
dampers and without additional uniform damping (i.e. with $\epsilon =0$).
For this problem the probability distribution function of scaled power
is given by Eq. (\ref{Py3}) and closed form expressions for the mean
value of $y$ and its variance are cumbersome. It is, therefore, reasonable
to carry out the corresponding integrations numerically. In contrast,
the naive averaging, which now also includes integration over the
resonance widths $\zeta _{r}$, distributed {[}1,8{]} according to
$\chi ^{2}$ distribution:\begin{equation*}
p\left( \frac{\zeta _{r}}{\overline{\Gamma }}\right) =\frac{M^{M}}{\Gamma \left( M\right) }\left( \frac{\zeta _{r}}{\overline{\Gamma }}\right) ^{M-1}\exp \left\{ -M\frac{\zeta _{r}}{\overline{\Gamma }}\right\} ,
\end{equation*}where $\overline{\Gamma }$ is mean resonance width, produces a relatively
compact answer for the statistics of $y$ (see Appendix B):\begin{equation}
\left\langle y\right\rangle \left(\pi \nu \right)^{2}=\left(\pi \nu \right)^{2}\frac{2}{\epsilon _{M}}\frac{M}{M-1},\label{ym}\end{equation}
\begin{align}
\left\langle y^{2}\right\rangle & =\frac{4M^{2}}{\left( M-2\right) \left(
M-1\right) \epsilon _{M}^{2}}f\left( \epsilon _{M},M\right)   \notag \\
&f\left( \epsilon _{M},M\right)  =2+\frac{4M\left( M-1\right) }{\left(
M-3\right) \epsilon _{M}} \label{ysm} \\
&-\frac{M\left( 4M\epsilon _{M}-4\epsilon
_{M}-M\right) }{\left( 2M-3\right) \epsilon _{M}^{2}}  \notag \\
& -\frac{\left( \epsilon _{M}+M\right) ^{4}}{M^{2}\left( 2M-3\right)
\epsilon _{M}^{2}}\left( 1+\frac{\epsilon _{M}}{M}\right) ^{4-2M},  \notag
\end{align}where $\epsilon _{M}=2\pi \nu N\overline{\Gamma }$. We note that
the $\chi ^{2}$ distribution for $\zeta _{r}$, is strictly correct
only for the case of $\overline{\Gamma }$ much less than mean level
distance. It does, however, correspond well with the actual distribution
{[}4{]} for more arbitrary value of $\overline{\Gamma }$, as long
as $M$ is large.

In order to compare Eqs. (\ref{ym}) and (\ref{ysm}) to corresponding
results obtained by numerical integration of Eq. (\ref{Py3}), it
is necessary to establish how $\overline{\Gamma }$ is related to
the parameters $N$, $M$ and $g$ of the supersymmetric calculation.
By definition, the mean scaled resonance width in open systems is
proportional to the product of modal density $\nu $ and average resonance
width $\overline{\Gamma }$ {[}4{]}. Under the condition of uniform
damping, the eigenmodes are equally damped, and the latter quantity
is just equal to the individual damper strength $\varepsilon $. In
general, the relationship is not that simple and is given by Moldauer-Simonius
formula {[}4{]}:\begin{eqnarray}
\overline{\gamma } & = & 2\pi \frac{\overline{\Gamma }}{\Delta }=2N\nu \pi \overline{\Gamma }=\frac{1}{2}\sum _{a}^{M}\ln \frac{g_{a}+1}{g_{a}-1},\notag \\
 &  & g_{a}=\frac{1}{2\pi \nu }\left(\frac{1}{\gamma _{a}}+\gamma _{a}\right),
\end{eqnarray}
for the mean scaled resonance width, where $\Delta =1/\nu N$ is the
mean eigenenergy spacing and $\gamma _{a}$ is the coupling constant
of $a$th channel. Note that our definitions of $\overline{\gamma }$
and eigenwidth $\zeta _{r}$ are different by factor of two from notation
of Ref. {[}4{]}. For the uniformly damped system we find:\begin{equation}
\overline{\gamma }\simeq \frac{1}{2}\sum _{a}^{N}\ln \frac{\frac{1}{2\pi \nu \varepsilon }+1}{\frac{1}{2\pi \nu \varepsilon }-1}\sim 2\pi \nu \varepsilon N\equiv \epsilon ,\end{equation}
which was also shown at the end of Section II. Thus, we conclude that
our parameter $\epsilon $ coincides with the mean scaled resonance
width in the limit of large number of equivalent weak channels. Moreover,
parameter $\epsilon _{M}$ in Eqs. (\ref{ym}), (\ref{ysm}) is the
same as $\overline{\gamma }$. %
\begin{figure*}[t]
\begin{center}\includegraphics{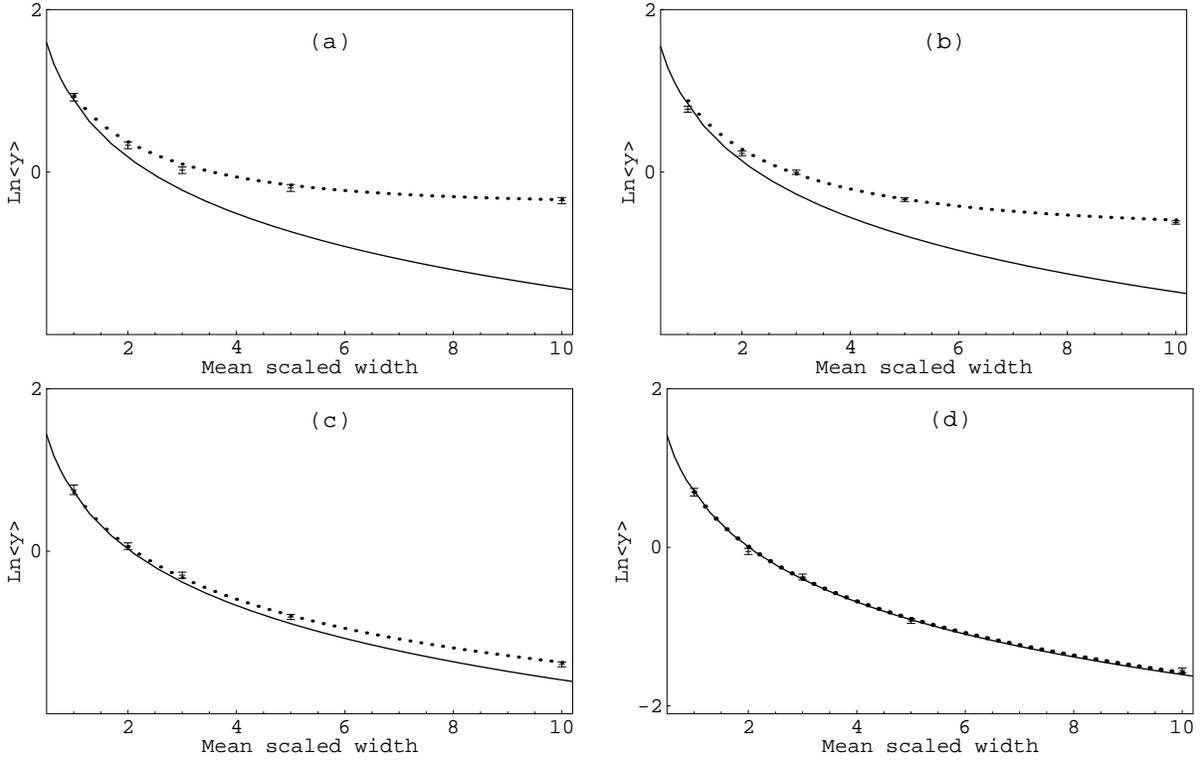}\end{center}

\caption{$Log\left\langle y\right\rangle $ plotted as a function of mean
scaled width $\overline{\gamma }$ for different number of channels:
(a) $M=6$; (b) $M=8$; (c) $M=40$; (d) $M=400$. Here we computed
$G_{ij}(E=0)$ ($i\neq j$, and $i>M,j>M$) for fixed $i$ and $j$
and averaged $y$ over $1500$ samples from the ensemble of $H$.
The naive averaging prediction (solid line) is compared to the prediction
by supersymmetry method (dotted line). The one sigma error bars were
computed based on the observed variance of $y$.}
\end{figure*}

\begin{figure*}[htbp]
\includegraphics{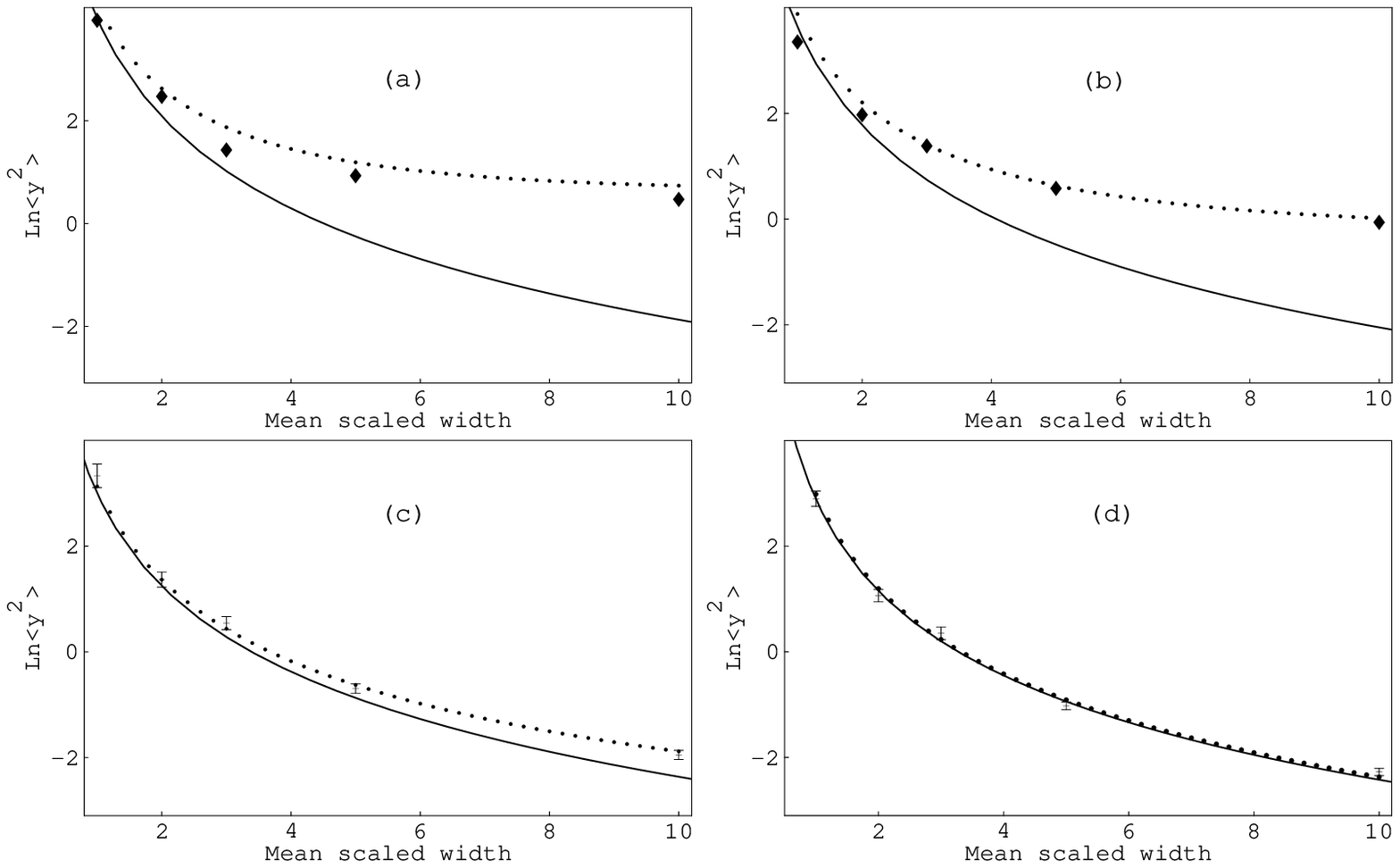}

\caption{$Log\left\langle y^{2}\right\rangle $ plotted as a function of mean
scaled width $\overline{\gamma }$for different number of channels:
(a) $M=6$; (b) $M=8$; (c) $M=40$; (d) $M=400$. $\left\langle y^{2}\right\rangle $
was obtained from the same data for $G_{ij}(E=0)$ as $\left\langle y\right\rangle $
in Fig. 4. The naive averaging prediction (solid line) is compared
to the prediction by supersymmetry method (dotted line). The error
bars in cases (c) and (d) were computed based on the observed variance
of $y^{2}$. Error bars for cases (a) and (b) are omitted, as one
sigma bars would misrepresent confidence intervals. Indeed, standard
deviation in case (a) do not exist.}
\end{figure*}

The averaging, performed in Ref. {[}12{]} for the uniformly damped
GOE system led to the dependence of the first two moments of the transmitted
ultrasonic power on a single structural parameter, the modal overlap
$\mathcal{M}$. $\mathcal{M}$ was defined in Ref. {[}12{]} in terms
of average imaginary part of the eigenfrequency $\omega _{r}$ and
modal density: $\mathcal{M}=2\pi \left\langle Im\omega _{r}\right\rangle \left(\partial N/\partial \omega \right)$.
We see that modal overlap may be identified with $\overline{\gamma }$. 

$\left\langle y\right\rangle $ and $\left\langle y^{2}\right\rangle $
as predicted by supersymmetric calculation (from Eq. (\ref{Py3}))\ and
by 'naive' averaging (Eqs. (\ref{ym}) and (\ref{ysm})) are compared
in Fig. 4 and Fig. 5 with numerical results for several different
values of $M$ and $\overline{\gamma }$. The prediction by the supersymmetry
method agrees with numerical results. In contrast, the results of
the 'naive' averaging underestimate both first and second moments
of the power, except for very large $M$, close to the uniform damping
case.

Finally, in connection with discussion of Ref. {[}12{]} we present
the comparison of the relative variance ($Relvar=\left\langle y^{2}\right\rangle /\left\langle y\right\rangle ^{2}-1$)
of power in Fig. 6, and compare the supersymmetric and naive predictions.%
\begin{figure*}[htbp]
\includegraphics{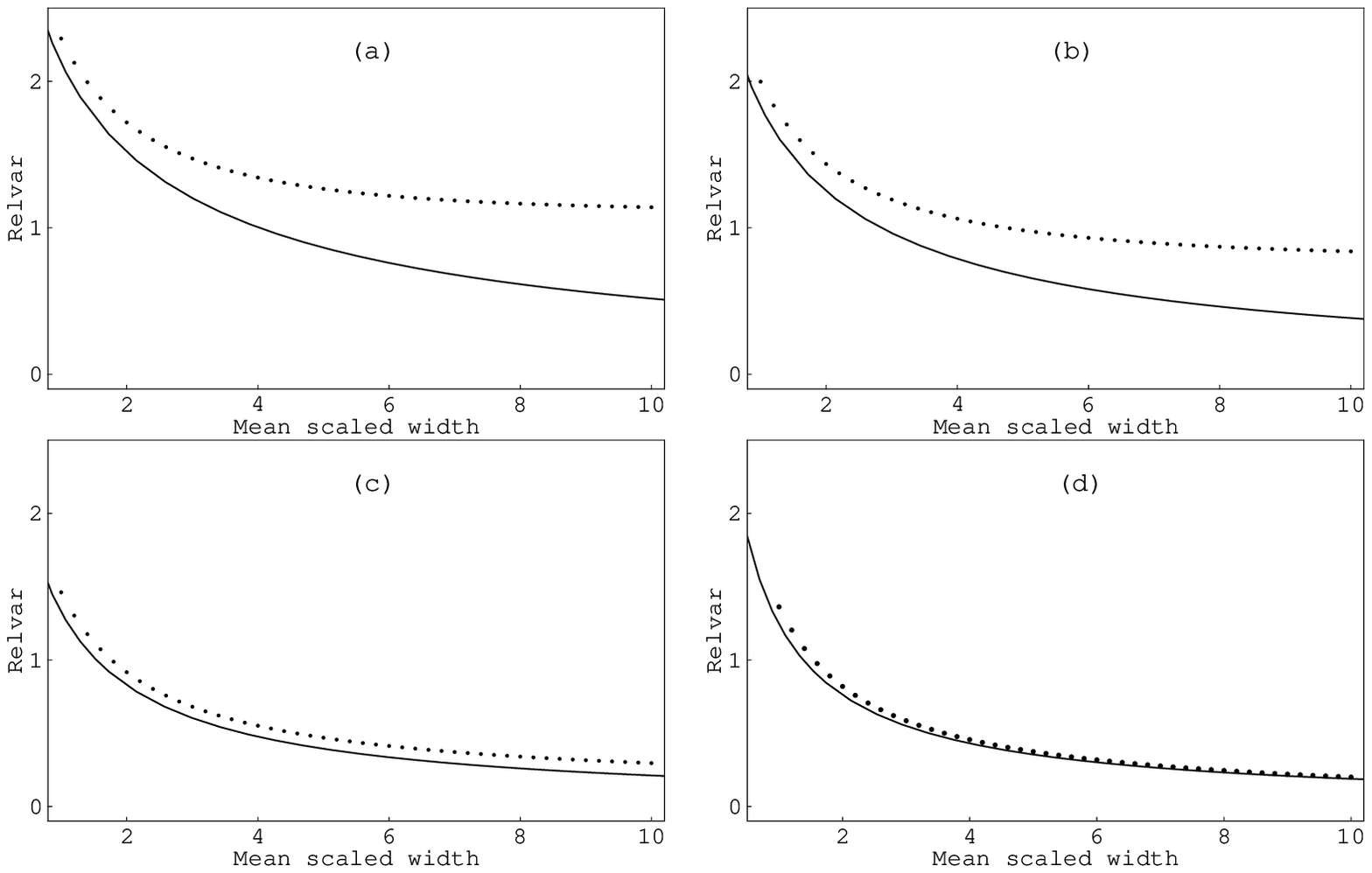}

\caption{Relative variance of power as a function of mean scaled width $\overline{\gamma }$
for different number of channels: (a) $M=6$; (b) $M=8$; (c) $M=40$;
(d) $M=400$ The results of naive averaging (solid line) consistently
underestimate the results by supersymmetry method (dotted line).}
\end{figure*}

\section{Conclusions}

We investigated the statistical behavior of the power transmitted
in a closed RMT system with internal dissipation, or an open RMT system
coupled to the exterior via a finite number of equally strong channels.
Using the supersymmetry method for systems with broken time reversal
invariance we derived an expression for the probability distribution
function for this quantity and studied its first two moments. The
theoretical predictions were compared to the results of numerical
simulations on GUE systems with dissipation, and to the results of
a 'naive' theory based on the RMT eigenstatistics of a closed non-dissipative
system. The results of the supersymmetric calculation agree with the
numerical data for the full range of parameters studied.

The naive averaging predictions are in general inconsistent with numerical
results, because its assumptions ($\chi ^{2}$ distribution of resonance
widths and decoupled uncorrelated Gaussian eigenmode amplitudes) follow
from the first order perturbation theory, valid for small scaled resonance
width. However, because the $\chi ^{2}$ distribution reduces to the
exact distribution for the case of uniform damping or in the limit
of a large number of weak channels, the naive theory is accurate in
this limit, for all values of scaled level width.

\begin{acknowledgments}
This work was supported by grants from the National Science Foundation
{[}CMS 99-88645 and CMS 0201346{]}, by computational resources from
the National Center of Supercomputing Applications, by EPSRC grant
GR/R13838/01 \char`\"{}Random matrices close to unitary or Hermitian\char`\"{}
and by Vice-Chancellor grant from Brunel University. IR would like
to thank International Programs in Engineering in University of Illinois
at Urbana-Champaign for financial support, the Department of Mathematical
Sciences at Brunel University for financial support and hospitality
during his visit and Antonio M. Garcia-Garcia for valuable suggestions. 
\end{acknowledgments}
\appendix

\section{Evaluation of the superintegral}

The result of 'Gaussian' integration in Eq. (\ref{ttnh}) has to be
simplified. To bring the integrand into a form convenient for saddle
point integration, we use the series of identities for the supermatrix
$f=L^{1/2}\widetilde{f}L^{1/2}$ ($L^{2}=I_{4}$) \begin{align}
\widetilde{f}& =-EI_{4}\otimes I_{N}-R\otimes I_{N}-i\Lambda \otimes \Gamma 
\notag \\
& =\left( I_{N}\otimes I_{4}-i\Gamma \otimes \left( \Lambda \mathfrak{G}^{-1}\right) \right) \mathfrak{G,}  \\
\widetilde{f}^{\,-1}& =\mathfrak{G}^{-1}\left( I_{4}\otimes I_{N}-\left(
\Lambda \mathfrak{G}^{-1}\right) \otimes i\Gamma \right) ^{-1}  \notag \\
& =\mathfrak{G}^{-1}\sum\limits_{k=0}^{\infty }\left( i\right) ^{k}\left[
\left( \Lambda \mathfrak{G}^{-1}\right) \otimes i\Gamma \right] ^{k},  \notag
\end{align}where $\mathfrak{G}=-EI_{4}-R$ was introduced. Supermatrix $\widetilde{f}^{\, -1}$
is diagonal in $i$ and $j$, thus:\begin{equation}
f^{bb}=L^{1/2}\widetilde{f}^{\, bb}L^{1/2},f_{12,\, bb_{ii}}f_{21,\, bb_{jj}}=\mathfrak{G}_{12,bb}\mathfrak{G}_{21,bb}.\notag \end{equation}

Substituting $Sdetf^{\, -1}=\exp \left\{ -\underset{i}{Str}\ln \widetilde{f}\right\} $
into the result of Gaussian integration with respect to the supervector
components, and considering\begin{align}
\left\langle T^{n}\right\rangle _{H}& =\left( n!\right) ^{-2}\int \left[ dR\right] \left( \mathfrak{G}_{\ 12,\,bb\,}^{-1}\mathfrak{G}_{\
21,bb}^{-1}\right) ^{n}S\det f^{\,-1}  \notag \\
& \times \exp \left\{ -\frac{N}{2}StrR^{2}+i\varepsilon NStrR\Lambda
\right\} ,   \label{a2} 
\end{align}we separated the terms in the exponent according to their order in
$N$ and obtained Eq. (\ref{Tn2}):\begin{align}
&\mathcal{L}\left[ R\right]  =\frac{1}{2}StrR^{2}+Str\ln \mathfrak{G,} 
\notag \\
&\delta \mathcal{L} =i\varepsilon NStrR\Lambda -Str\ln \left[ I_{N}-i\Gamma
\otimes \left( \Lambda \mathfrak{G}^{-1}\right) \right]    \notag \\
&= i\varepsilon NStrR\Lambda -MStr\ln \left[ I_{4}-i\gamma \Lambda \mathfrak{G}^{-1}\right] .   \label{a3}
\end{align}The last identity was proved by expanding the logarithm into the series
(see Ref. {[}4{]}). 

After the Gaussian integration around the saddle point in Eq. (\ref{a2})
the probability distribution function for the scaled power is expressed
as an integral over the manifold formed by supermarices $Q=\mathfrak{T}^{-1}\Lambda \mathfrak{T}$:\begin{equation*}
\mathcal{P}\left( y\right) =\int \left[ dQ\right] \delta \left(
y-Q_{12,bb}Q_{21,bb}\right) Y\left( Q\right) ,
\end{equation*}$Q$ is parametrized by four commuting variables $\lambda _{1}$,
$\lambda _{2}$, $\mu _{1}$, $\mu _{2}$ and four anticommuting $\alpha $,
$\alpha ^{\ast }$, $\beta $, $\beta ^{\ast }$ \cite{key-4,key-17}.
Here $\lambda _{1}\in \left(1,\infty \right)$, $\lambda _{2}\in \left(-1,1\right)$,
and $\left|\mu _{1}\right|^{2}=\lambda _{1}^{2}-1$, $\left|\mu _{2}\right|^{2}=1-\lambda _{2}^{2}$.
We can also introduce another set of variables, according to $\lambda _{1}=\cosh \theta _{1}$,
$\mu _{1}=\sinh \theta _{1}\exp \left\{ i\phi _{1}\right\} $, $\lambda _{2}=\cos \theta _{2}$,
$\mu _{2}=\sin \theta _{2}\exp \left\{ i\phi _{2}\right\} $, where
$\theta _{1}\in \left(0,\infty \right)$, $\theta _{2}\in \left(0,\pi \right)$,
$\phi _{1}$, $\phi _{2}\in \left(0,2\pi \right)$. Next we observe
{[}4{]}\begin{widetext}

\begin{align}
&StrQ\Lambda  =-2i\left( \lambda _{1}-\lambda _{2}\right) ,  Y\left( Q\right) =S\det \,^{-M}\left[ I_{4}+i\frac{E}{2}\gamma \Lambda
+i\pi \nu \gamma Q\Lambda \right]  \exp \left\{ -i\varepsilon \pi \nu
NStrQ\Lambda \right\}    \label{a4} \\
=&\left( \frac{1+2\pi \nu \gamma \lambda _{2}+\gamma ^{2}}{1+2\pi \nu
\gamma \lambda _{1}+\gamma ^{2}}\right) ^{M}\exp \left\{ -2\varepsilon \pi
\nu NStr\left( \lambda _{1}-\lambda _{2}\right) \right\}  =\left( \frac{g+\lambda _{2}}{g+\lambda _{1}}\right) ^{M}\exp \left\{
-2\varepsilon \pi \nu NStr\left( \lambda _{1}-\lambda _{2}\right) \right\} ,
\notag
\end{align}

\end{widetext}where $g=\left(1/\gamma +\gamma \right)/\left(2\pi \nu \right)\, \, $,\begin{align*}&
Q_{12,bb} =\mu _{1}\left( 1-\alpha ^{\ast }\alpha /2\right) \left( 1+\beta
^{\ast }\beta /2\right) -\alpha ^{\ast }\beta \mu _{2}^{\ast },  \notag \\
&Q_{21,bb}=\mu _{1}^{\ast }\left( 1-\alpha ^{\ast }\alpha /2\right) \left(
1+\beta ^{\ast }\beta /2\right) +\alpha \beta ^{\ast }\mu _{2},  
 \notag
\end{align*}\begin{align}
 &Q_{12,bb}Q_{21,bb} =\left\vert \mu _{1}\right\vert ^{2}+\left\vert \mu
_{1}\right\vert ^{2}\alpha ^{\ast }\beta ^{\ast }\alpha \beta \notag \\
&+\left\vert
\mu _{2}\right\vert ^{2}\alpha ^{\ast }\beta ^{\ast }\alpha \beta  \notag 
 +\left\vert \mu _{1}\right\vert ^{2}\left( \beta ^{\ast }\beta -\alpha
^{\ast }\alpha \right) \notag  \\
&+\left\vert \mu _{1}\right\vert \left\vert \mu
_{2}\right\vert e^{i\left( \phi _{1}+\phi _{2}\right) }\alpha \beta ^{\ast }
\label{a5} \\
&-\left\vert \mu _{1}\right\vert \left\vert \mu _{2}\right\vert e^{-i\left(
\phi _{1}+\phi _{2}\right) }\alpha ^{\ast }\beta ,  \notag
\end{align}and the integration measure was defined as\begin{equation*}
dQ=\frac{d\alpha ^{\ast }d\beta ^{\ast }d\alpha d\beta d\lambda _{1}d\lambda
_{2}d\phi _{1}d\phi _{2}}{\left( 2\pi \left( \lambda _{1}-\lambda
_{2}\right) \right) ^{2}}.
\end{equation*}Substituting Eqs. (\ref{a4}), (\ref{a5}) we proceed with integration
with respect to Grassman variables. First, we need to expand the delta
function retaining only the terms of zero and maximum order in these
variables \cite{key-4,key-17,key-18}. Setting $Q_{12,bb}Q_{21,bb}=-1+\lambda _{1}^{2}+z$
we expand the delta function,\begin{gather*}
\delta \left( y-Q_{12}^{bb}Q_{21}^{bb}\right) =\delta \left( y+1-\lambda
_{1}^{2}-z\right) =\delta \left( y+1-\lambda _{1}^{2}\right)  \\
-\left( \delta _{z}^{\prime }\left( y+1-\lambda _{1}^{2}\right) +\delta
_{zz}^{\prime \prime }\left( y+1-\lambda _{1}^{2}\right) \left( 1-\lambda
_{1}^{2}\right) \right)  \\
\times \left( \lambda _{1}^{2}-\lambda _{2}^{2}\right) \alpha ^{\ast }\beta
^{\ast }\alpha \beta +...=\delta \left( y+1-\lambda _{1}^{2}\right)  \\
+\left( \frac{d}{dy}+y\frac{d^{2}}{dy^{2}}\right) \delta \left( y+1-\lambda
_{1}^{2}\right) \alpha ^{\ast }\beta ^{\ast }\alpha \beta +...,
\end{gather*}where we used the fact that the argument of delta function is linear
in $y$ ($z$ and $1-\lambda _{1}^{2}$), in order to be able to take
the differential operator out of the integral. Then, we calculate
the integral over $\phi _{1},\phi _{2}$ and Grassman variables,\begin{align}
\mathcal{P}\left( y\right) & =\delta \left( y\right) +\left( \frac{d}{dy}+y\frac{d^{2}}{dy^{2}}\right)\notag \\
\times&\int_{1}^{\infty }d\lambda
_{1}\int_{-1}^{1}d\lambda _{2}\delta \left( y+1-\lambda _{1}^{2}\right)  
 \\
\times & \frac{\lambda _{1}^{2}-\lambda _{2}^{2}}{\left( \lambda
_{1}-\lambda _{2}\right) ^{2}}\exp \left\{ -\epsilon \left( \lambda
_{1}-\lambda _{2}\right) \right\} \left( \frac{g+\lambda _{2}}{g+\lambda _{1}}\right) ^{M},  \notag
\end{align}where $\epsilon =2\varepsilon \pi \nu N$ and we have used\begin{align*}
\int& \left[ dQ\right] \exp \left\{ -\epsilon \left( \lambda _{1}-\lambda
_{2}\right) \right\} \notag \\
\times &\left( \frac{g+\lambda _{2}}{g+\lambda _{1}}\right) 
^{M}  \delta \left( y+1-\lambda _{1}^{2}\right) =\delta \left( y\right) ,
\end{align*}for the 'Efetov-Wegner' term {[}4,18{]}. Integration with respect
to $\lambda _{1}$ is simple because of the presence of the delta
function.\begin{align*}
\mathcal{P}\left( y\right) & =\delta \left( y\right) +\left( \frac{d}{dy}+y\frac{d^{2}}{dy^{2}}\right) \frac{\exp \left\{ -\epsilon \sqrt{1+y}\right\}
\theta \left( y\right) }{2\sqrt{1+y}\left( g+\sqrt{1+y}\right) ^{M}} \\
& \times \int_{-1}^{1}d\lambda _{2}\frac{\sqrt{1+y}+\lambda _{2}}{\sqrt{1+y}-\lambda _{2}}\exp \left\{ \epsilon \lambda _{2}\right\} \left( g+\lambda
_{2}\right) ^{M} \\
& =\delta \left( y\right) +\left( \frac{d}{dy}+y\frac{d^{2}}{dy^{2}}\right)
\theta \left( y\right) \mathfrak{F}(y).
\end{align*}\bigskip Using property $\delta \left(y\right)=-y\delta ^{\prime }\left(y\right)$,
we arrive at:\begin{eqnarray}
\mathcal{P}\left(y\right) & =\delta \left(y\right)+\delta \left(y\right)y\frac{d}{dy}\mathfrak{F}(y)\notag  & \\
 & +\left(\frac{d}{dy}+y\frac{d^{2}}{dy^{2}}\right)\mathfrak{F}(y), & \label{a7}
\end{eqnarray}
Eq. (\ref{a7}) completes the calculation of probability distribution
function for the scaled power. This equation yields Eq. (\ref{Py2})
upon substitution of: $\lim _{y\rightarrow 0}y\frac{d}{dy}\mathfrak{F}(y)=-1$. 

Finally, we set $\epsilon =0$ and derive Eq. (\ref{Py3}). We note
that integral in $\mathfrak{F}(y)$ is a table integral:\begin{align*}
\mathcal{P}\left( y\right) & =\left( \frac{d}{dy}+y\frac{d^{2}}{dy^{2}}\right)  \\
& \times \frac{\left( g+1\right) ^{M+1}\,f\left( g+1\right) \,-\left(
g-1\right) ^{M+1}f\left( g-1\right) }{\left( M+1\right) \left( g+\sqrt{y+1}\right) ^{M+1}}, \\
f\left( u\right) & =_{2}F_{1}\left( M+1,1,M+2,\frac{u}{g+\sqrt{y+1}}\right) .
\end{align*}Thus, it is possible to apply differential operator to obtain the
final form of $\mathcal{P}\left(y\right)$ for this case. However,
we notice: \begin{align}
\mathfrak{F}(y)=&\frac{-1}{\left( g+\sqrt{y+1}\right) ^{M}}\int_{-1}^{1}\frac{\left( g+\lambda _{2}\right) ^{M}}{\lambda _{2}-\sqrt{y+1}}d\lambda _{2} 
 \label{a8} \\
-&\frac{1}{2\sqrt{y+1}\left( g+\sqrt{y+1}\right) ^{M}}\int_{-1}^{1}\left(
g+\lambda _{2}\right) ^{M}d\lambda _{2}.  \notag
\end{align}

The second term in the above equation can be evaluated immediately,
while for the first one we can use an identity:

\begin{align*}
\left( g+\lambda _{2}\right) ^{M}=&\sum\limits_{m=0}^{M}\left( g+\sqrt{y+1}\right) ^{m} \notag \\
\times&\left( \lambda _{2}-\sqrt{y+1}\right) ^{M-m}  \frac{M!}{m!\left(
M-m\right) !}
\end{align*}

We integrate each term in Eq. (\ref{a8}) separately, and after the
first differentiation of the result with respect to $y$, the series
can be summed back, so that the remaining procedure becomes straightforward
and leads to Eq. (\ref{Py3}).

\section{Moments calculation for the case of $M$ equivalent dampers}

In this Appendix we demonstrate the intermediate steps leading to
the Eqs. (\ref{ysb1}), (\ref{ym}), (\ref{ysm}). We start with the
modal expansion for $T^{2}$ without making an assumption about uniform
damping:\begin{align}
y^{2}\left( \pi \nu \right) ^{4}& =\sum\limits_{r,m,l,k}\frac{u_{i}^{r}u_{j}^{r\ast }}{E-E_{r}-i\zeta _{r}}\frac{u_{i}^{m\ast }u_{j}^{m}}{E-E_{m}+i\gamma _{m}} 
 \\
& \times \frac{u_{i}^{l\ast }u_{j}^{l}}{E-E_{l}-i\zeta _{l}}\frac{u_{i}^{k\ast }u_{j}^{k}}{E-E_{k}+i\gamma _{k}}.  \notag
\end{align}

Absence of correlation between different eigenmodes produces the following
result for the variance of $y$:\begin{widetext}

\begin{align}
\left\langle y^{2}\right\rangle \left( \pi \nu \right) ^{4}& =\sum_{r}\frac{\left\langle \left\vert u_{i}^{r}\right\vert ^{4}\right\rangle \left\langle
\left\vert u_{j}^{r}\right\vert ^{4}\right\rangle }{\left( E-E_{r}-i\zeta
_{r}\right) ^{2}\left( E-E_{r}+i\zeta _{r}\right) ^{2}} +\sum\limits_{r\neq l}\frac{\left\langle \left\vert u_{i}^{r}\right\vert
^{2}\right\rangle \left\langle \left\vert u_{j}^{r}\right\vert
^{2}\right\rangle \left\langle \left\vert u_{i}^{l}\right\vert
^{2}\right\rangle \left\langle \left\vert u_{j}^{l}\right\vert
^{2}\right\rangle }{\left( E-E_{r}-i\zeta _{r}\right) ^{2}\left(
E-E_{l}+i\zeta _{l}\right) ^{2}}  \label{b2} \notag \\
& +\sum\limits_{r\neq l}\frac{\left\langle \left\vert u_{i}^{r}\right\vert
^{2}\right\rangle \left\langle \left\vert u_{j}^{r}\right\vert
^{2}\right\rangle \left\langle \left\vert u_{i}^{l}\right\vert
^{2}\right\rangle \left\langle \left\vert u_{j}^{l}\right\vert
^{2}\right\rangle }{\left( E-E_{r}-i\zeta _{r}\right) \left( E-E_{r}+i\zeta
_{r}\right) \left( E-E_{l}-i\zeta _{l}\right) \left( E-E_{l}+i\zeta
_{l}\right) }  
\end{align}

\end{widetext}

Next, we replace summation over $E_{r}$ and $E_{l}$ with integration
($\sum _{r}\rightarrow N\nu \int dE_{r}$) and take into account the
correlation between the GUE eigenvalues in Eq. (\ref{b2}) by introducing
the factor $1-Y_{2}\left(\pi N\nu \left(E_{r}-E_{l}\right)\right)$,
{[}1{]}:\begin{widetext}

\begin{align}
\left\langle y^{2}\right\rangle & =\frac{N\nu \left\langle \left\vert
u\right\vert ^{4}\right\rangle ^{2}}{\left( \pi \nu \right) ^{4}}\int_{-\infty }^{\infty }\frac{dx}{\left( x-i\zeta _{r}\right) ^{2}\left(
x+i\zeta _{r}\right) ^{2}}+\frac{2\left( N\nu \right) ^{2}\left\langle
\left\vert u\right\vert ^{2}\right\rangle ^{4}}{\left( \pi \nu \right) ^{4}}
\label{b3} \\
& \times \int_{-\infty }^{\infty }\int_{-\infty }^{\infty }\frac{\left(
1-Y_{2}\left( \pi N\nu z\right) \right) dxdz}{\left( x^{2}+\zeta _{r}\right)
\left( \left( x-z\right) ^{2}+\zeta _{l}^{2}\right) }+\frac{\left( N\nu
\right) ^{2}\left\langle \left\vert u\right\vert ^{2}\right\rangle ^{4}}{\left( \pi \nu \right) ^{4}} \int_{-\infty }^{\infty }\int_{-\infty }^{\infty }\frac{\left(
1-Y_{2}\left( \pi N\nu z\right) \right) dxdz}{\left( x-i\zeta _{r}\right)
^{2}\left( x-z+i\zeta _{l}\right) ^{2}},  \notag
\end{align}

\end{widetext}where $x=$\bigskip $E-E_{r}$, $z=E_{r}-E_{l}$ and the Dyson two-level
correlation function for the GUE is $Y_{2}\left(\xi \right)=\left(\sin \xi /\xi \right)^{2}$.
Integration over $x$ and $z$ in Eq. (\ref{b3}) for the case of
uniform damping $\zeta _{r}=\zeta _{l}=\varepsilon $ yields Eq. (\ref{ysb1}):\begin{align}
&\left\langle y^{2}\right\rangle  =\frac{N\nu \left\langle \left\vert
u\right\vert ^{4}\right\rangle ^{2}}{\left( \pi \nu \right) ^{4}}\frac{\pi }{2\zeta _{r}^{3}}+\frac{2\left( N\nu \right) ^{2}\left\langle \left\vert
u\right\vert ^{2}\right\rangle ^{4}}{\left( \pi \nu \right) ^{4}}  \notag \\
& \times \frac{\zeta _{r}+\zeta _{l}}{2\zeta _{r}\zeta _{l}}\int_{-\infty
}^{\infty }\frac{\left( 1-Y_{2}\left( \pi N\nu z\right) \right) dz}{z^{2}+\left( \zeta _{r}+\zeta _{l}\right) ^{2}}  \\
& =\left\langle \left\vert u\right\vert ^{4}\right\rangle ^{2}\frac{N}{2\left( \pi \nu \right) ^{3}\varepsilon ^{3}}+\left\langle \left\vert
u\right\vert ^{2}\right\rangle ^{4}\frac{1}{4\left( \pi \nu \right)
^{4}\varepsilon ^{4}}  \notag \\
& \times \left[ 1+8\left( N\nu \right) ^{2}\pi ^{2}\varepsilon ^{2}-\exp
\left\{ -4N\nu \pi \varepsilon \right\} -4N\nu \pi \varepsilon \right] , 
\notag
\end{align}upon substitution $\left\langle \left|u\right|^{4}\right\rangle /\left\langle \left|u\right|^{2}\right\rangle ^{2}=2$,
(as is the case for complex Gaussian random numbers) and $\left\langle \left|u\right|^{2}\right\rangle =1/N$. 

In the case of finite number $M$ of weak dampers ($\epsilon =0$)
the ensemble averaging includes an integration with over a distribution
of widths, given by {[}1,8{]}:\begin{equation}
p\left(\frac{\zeta _{r}}{\overline{\Gamma }}\right)=\frac{M^{M}}{\Gamma \left(M\right)}\left(\frac{\zeta _{r}}{\overline{\Gamma }}\right)^{M-1}\exp \left\{ -M\frac{\zeta _{r}}{\overline{\Gamma }}\right\} ,\end{equation}
where $\overline{\Gamma }$ is average resonance width and $\Gamma \left(M\right)$
is a Gamma function. Starting with Eq. (\ref{Pofy1}) we average $y$
over the eigenmodes and eigenenergies to get:\begin{equation}
\left\langle y\right\rangle \left(\pi \nu \right)^{2}=N\nu \left\langle \left|u\right|^{2}\right\rangle ^{2}\frac{\pi }{\zeta _{r}}.\end{equation}
Which becomes Eq. (\ref{ym}) upon integration over $p\left(\zeta _{r}\right)$.
We indicate this averaging by overbar:\begin{equation*}
\left\langle y\right\rangle \left( \pi \nu \right) ^{2}=\overline{N\nu
\left\langle \left\vert u\right\vert ^{2}\right\rangle ^{2}\frac{\pi }{\zeta
_{r}}}=\left\langle \left\vert u\right\vert ^{2}\right\rangle ^{2}\frac{\pi
N\nu }{\overline{\Gamma }}\frac{M}{M-1}.
\end{equation*}To evaluate the second moment of power we integrate over $x$, and
then over $\zeta _{r}$ and $\zeta _{l}$ in Eq. (\ref{b3}):\begin{align}
\left\langle y^{2}\right\rangle \left(\pi \nu \right)^{4} & =\overline{N\nu \left\langle \left|u\right|^{4}\right\rangle ^{2}\frac{\pi }{\zeta _{r}^{3}}}+\overline{2\left(N\nu \right)^{2}\left\langle \left|u\right|^{2}\right\rangle ^{4}I}\notag \label{b7}\\
I= & \frac{\zeta _{r}+\zeta _{l}}{2\zeta _{r}\zeta _{l}}\int _{-\infty }^{\infty }\frac{\left(1-Y_{2}\left(\pi N\nu z\right)\right)dz}{z^{2}+\left(\zeta _{r}+\zeta _{l}\right)^{2}}.
\end{align}

The remaining integral with respect to $z$ in $\overline{I}$ it
is convenient to use the Fourier transform of $Y_{2}\left(\xi \right)$,
which has a form:\begin{align*}
b\left( q\right) & =\int_{-\infty }^{\infty }Y_{2}\left( \xi \right) \exp
\left\{ 2\pi i\xi q\right\} d\xi =1-\left\vert q\right\vert ,\left\vert
q\right\vert \leq 1, \\
b\left( q\right) & =0,\left\vert q\right\vert \geq 1,
\end{align*}

\begin{align}
\overline{I}= & \\
= & \overline{\frac{\zeta _{r}+\zeta _{l}}{4\zeta _{r}\zeta _{l}}\int _{-1}^{1}\left(1-\frac{\left|q\right|}{2\pi }\right)\exp \left\{ -N\nu \left(\zeta _{r}+\zeta _{l}\right)\right\} dq}.\notag 
\end{align}

The average over $\zeta _{r}$ and $\zeta _{l}$ is now straightforward.
Finally, substituting the result into the Eq.(\ref{b7}), and taking
into account $\left\langle \left|u\right|^{4}\right\rangle /\left\langle \left|u\right|^{2}\right\rangle ^{2}=2$,
we obtain the second moment in its closed form (Eq. (\ref{ysm})).
The derivation presented above assumes that resonance widths $\zeta _{r}$
and eigenmodes $u^{r}$ are statistically independent.

\end{document}